\documentstyle[prl,aps]{revtex}

\begin{document}

\draft
\twocolumn

\title{ Can Quantum Computer Perform Better than Classical ?}

\author{Robert Alicki}

\address{Institute of Theoretical Physics and Astrophysics, University
of Gda\'nsk, Wita Stwosza 57, PL 80-952 Gda\'nsk, Poland}

\date{\today}
\maketitle

\begin{abstract}
We present a theoretical  model of a quantum device which can
factorize any number $N$ in two steps, by preparing an input state and
then
performing a proper measurement process. However, the analysis reveals
that the duration
of state preparation and measurement is at least proportional to $N$ and
hence
the computation is not efficient. On the other hand, the energy
consumption of this quantum computer grows like $\log N$ while for
classical ones is exponential
in the input bit size $m= \log_2N$. These results suggest the existence
of a generalized Heisenberg relation which put limits on the efficiency
of quantum computers in terms of the total computation time, the total
energy consumption and the classical complexity of the problem.
\end{abstract}


The action of a quantum computer is described on the abstract level
in the following way. We have a quantum system with $N$ orthogonal
states
(computational basis) which can store  $n= \log_2 N $ bits of
information.
Firstly, we have to prepare our quantum system in an initial (input)
state.
Then a quantum algorithm is performed which is realized as a sequence
of $T$ unitary transformation called quantum gates. Finally, the output
state is measured. The number $T$ of involved quantum gates
is assumed to be proportional to the physical time needed to achieve a
given
task. The algorithm is called efficient if $T$ is polynomial
in $m = \log_2 N$. The celebrated Shor's quantum algorithm [1] which
factorizes numbers into primes is efficient while its classical
counterparts
are not. This remarkable result had an enormous impact on the
development
of the whole field of quantum information and quantum computing [2].
\par
There are, however, the following factors which can spoil the
performance of
a quantum computer:
\par
A) The unavoidable inaccuracies in manufacturing of a hardware
which produce deviations of the real computer's Hamiltonian from the
designed one.
 \par
B) The number of elementary transformations which form a quantum
algorithm is by no means unique and does not determine the real physical

time of computation. In particular, by increasing the energy level
spacing of
the corresponding computer's Hamiltonian we can speed up its time
evolution.
On the other hand the decoherence effects due to an interaction with
an environment typically grow
in a nonlinear way with the energy level spacing.
\par
 C) A finite duration of the input state preparation procedure and
the output state measurement process should be taken into account.
In particular, one can expect that the Heisenberg energy-time
uncertainty
relation may put here some universal limits.
\par
The factor A) poses a rather technological problem and will be not
discussed here. The questions raised in B) will be discussed in details
in the forthcomming publication [3].
Our goal is to show by constructing an explicit but purely theoretical
model that there exist fundamental obstacles related to C) which could
destroy the efficiency of quantum algorithms.
\par
We describe the operation of a fictitious quantum machine which
can factorize any number $N$  with $T=0$ quantum gates,
it means by preparing an input state and
a proper measurement of it only. Although formally, there are no quantum
gates in
this model the quantum algorithm exists in the form of a specially
designed
Hamiltonian of the system. The dynamics governed by the system's
Hamiltonian
influences state preparation and measurements processes. In this case
the problems raised in B) are neglected.
\par
Consider a resonant cavity which supports the states
of a photon (radiation modes)  with the frequencies being the logarithms

of prime numbers times a fixed frequency unit $\omega$
$$
\omega_q = \omega\log q\ ,\  q = 2,3,5,7,11,13,...
\eqno(1)
$$
The second quantization Hamiltonian of the electromagnetic field in the
cavity
written in terms of annihilation and creation operators ${\hat a}_q,
{\hat a}^+_q$
$$
{\hat H} = \hbar \omega\sum_q (\log q)\  {\hat a}^+_q {\hat a}_q
\eqno(2)
$$
possesses nondegenerated eigenvalues being proportional to the
logarithms
of all natural numbers
$$
{\hat H}\psi_N = E_N \psi_N\ ,\ E_N = \hbar\omega\log N \ ,\  N =
1,2,3,4,...
\eqno(3)
$$
The structure of the corresponding eigenstates reveals the factorization

of $N$ into prime numbers. Namely,
$$
\psi_N \sim ({\hat a}^+_{q_1})^{m_1}({\hat a}^+_{q_2})^{m_2}\cdots
({\hat a}^+_{q_r})^{m_r}\psi_1
\eqno(4)
$$
where
$$
N = (q_1)^{m_1}(q_2)^{m_2}\cdots(q_r)^{m_r}
\eqno(5)
$$
and $\psi_1$ is a vacuum state. In words, at the state $\psi_N$ we have
$m_1$ photons of the frequency $\omega\log q_1$, $m_2$ photons of the
frequency
$\omega\log q_2$, ..., and $m_r$ photons of the frequency $\omega\log
q_r$.
Therefore, in principle one can find the factorization of any number $N$
in two steps. First we prepare the system in the state $\psi_N$ of a
given energy $E_N =\hbar\omega\log N$ by transferring this portion of
energy into the empty cavity. Then we open the cavity and perform a
spectral analysis
of the corresponding radiation field counting photons in different
modes.
\par
Let us discuss the possible preparation process. We perturb our quantum
system
being initially in a vacuum state by a weak external interaction
Hamiltonian ${\hat V}(t)$
which is periodic in time with a tunable frequency $\Omega$
$$
{\hat V}(t) = {\hat W} \cos (\Omega t)\ .
\eqno(6)
$$
Instead of a usual electromagnetic interaction linear in field we
assume that ${\hat W}$ is a sufficiently nonlinear function of
electromagnetic field operators which allows multiphoton excitations
such that the (virtual) transitions between the vacuum $\psi_1$ and any
state $\psi_N$ are possible i.e.
$$
<\psi_1,{\hat W} \psi_N> \neq 0\ ,\ for\ all\ N= 2 ,3,4,...
\eqno(7)
$$
For a given number $N$ which we want to factorize we tune the frequency
$\Omega$ to the value $\omega \log N$. The time dependent first-order
perturbation calculus [4] gives us the
probability of excitation of the state $\psi_M$
$$
p_M(t)= {2\over \hbar^2} |<\psi_1,{\hat W} \psi_M>|^2\ {\sin^2
\bigl\{{1\over 2}
\omega\bigl(\log M -\log N\bigr) t\bigr\}\over
\omega^2\bigl(\log M - \log N\bigr)^2}\ .
\eqno(8)
$$
As the energy level spacing around $E_N = \hbar\omega
\log N$  is $\delta E_N \approx
\hbar\omega/N$ it follows from the formula (8) that we have to wait
for a time at least of the order
$$
t \approx N \omega^{-1}
\eqno(9)
$$
to be sure that the selected state $\psi_N$ has
been prepared with much larger probability than the other neighboring
states
$\psi_M$. The similar estimation can be easily obtained for the duration

of the measurement process.
Therefore the total computation time $t_c$ grows exponentially with
$\log N$ like in the classical situation.
\par
It is obvious that the result obtained above can be treated as a special
case
of the Heisenberg time-energy uncertainty relation [4]
$$
\Delta t\cdot \Delta E \geq {\hbar\over 2}\ .
\eqno(10)
$$
Indeed in order to identify the energy of a quantum state with
an accuracy $\hbar\omega/N$ we need a time longer than $N/\omega$.
\par
One should notice, however, that our quantum computer is superior
to classical ones in respect of energy consumption, at least for the
case of
existing irreversible computers (see [5] for the theory of reversible
computations).
The energy
used for the factorization of $N$ is equal to $E_c = \hbar\omega
\log N$ while for the classical irreversible computers any logical step
consumes
an energy portion and hence the total energy cost of factorization
grows exponentially with the input bit size $m = \log_2 N$.
Taking into account the eq.(9) we obtain the following inequality
independent of an arbitrary frequency scale $\omega$
$$
t_c E_c >> \hbar N \log N\ .
\eqno(11)
$$
The form of the inequality (11) suggests the following general
hypothesis. There exists an inequality which puts universal limits on
the performance of a quantum computer in terms of the total computation
time $t_c$, the total energy consumption $E_c$
and the complexity ${\cal C}(\log_2 N)$ of the problem to be solved.
This "generalized Heisenberg relation" reads
$$
t_c E_c >> \hbar {\cal C} (\log_2 N)\ .
\eqno(12)
$$
The complexity ${\cal C}(\log_2 N)$ is a function of the input
bit size and is defined by a minimal number of logical steps needed
to solve the problem. The inequality (12) means that for a non-efficient
optimal classical algorithm the quantum computation
is also not efficient either with respect to the computation
time or the used energy.
\par
In order to prove
this hypothesis we cannot restrict ourselves to counting
quantum gates in the algorithm but we have to discuss physical
implementation of all steps of quantum computing including state
preparation and measurement processes.

\acknowledgments
The author thanks Micha\l\ , Pawe\l\ , and Ryszard Horodecki's and S.
Kryszewski for discussions.The work is supported by the Grant KBN
PB/273/PO3/99/16.

\end{document}